# Chromatic X-Ray imaging with a fine pitch CdTe sensor coupled to a large area photon counting pixel ASIC

**R. Bellazzini**[a,b]**, G. Spandre**[a*]**, A. Brez**[a]**, M. Minuti**[a]**, M. Pinchera**[a] **and P. Mozzo**[b]

[a] INFN Pisa
L.go Pontecorvo 3, 56127 Pisa, Italy
[b] PIXIRAD Imaging Counters s.r.l.
c/o INFN Pisa
*E-mail*: gloria.spandre@pi.infn.it

ABSTRACT: An innovative X-ray imaging sensor with intrinsic digital characteristics is presented. It is based on Chromatic Photon Counting technology. The detector is able to count individually the incident X-ray photons and to separate them according to their energy (two 'color' images per exposure). The energy selection occurs in real time and at radiographic imaging speed (GHz global counting rate). Photon counting, color mode and a very high spatial resolution (more than 10 l.p./mm at MTF50) allow to obtain an optimal ratio between image quality and absorbed dose. The individual block of the imaging system is a two-side buttable semiconductor radiation detector made of a thin pixellated CdTe crystal (the sensor) coupled to a large area VLSI CMOS pixel ASIC. 1, 2, 4, 8 tile units have been built. The 8 tiles unit has 25×2.5 cm² sensitive area. Results and images obtained from in depth testing of several configurations of the system are presented. The X-Ray imaging system is the technological platform of PIXIRAD Imaging Counters s.r.l., a recently constituted INFN spin-off company.

---

[*] Corresponding author.gloria.spandre@pi.infn.it

**Contents**

## 1. Introduction

The continuous progress and scale reduction of the CMOS technology has allowed the realization of pixellated VLSI ASICs with direct integration in each pixel of electronics of increased complexity and functionality [1,2,3]. By coupling a custom pixellated ASIC with a thin pixellated CdTe crystal, we have realized an innovative X-ray imaging sensor (PIXIRAD) based on Chromatic Photon Counting technology. The detector is able to count and separate in energy the X-ray photons transmitted through the object and converted in each pixel of the CdTe sensor to produce two '*color*' images of the object in a single exposure.

In contrast with the most traditional integrating technologies, the approach we have followed let to process, digitize and store each photon count in real time at the level of the single pixel.

Respect to the Amorphous Silicon Flat Panel, which is currently the state of the art technology, the Chromatic Photon Counting has a level of electronic noise of the elementary cell, extremely lower. This is consequence of the fact that each pixel of the sensor is directly connected to its complete electronic chain, which is therefore distributed over the entire surface of the sensor and not only at the periphery. This allows photon by photon processing to reduce the noise to the quantum level only.

## 2. The X-ray imaging sensor

The individual block of the PIXIRAD imaging system is two opposite side buttable and consists of a solid-state sensor (CdTe) connected to a CMOS readout ASIC by *flip-chip bonding* (bump bonding) technique. The system therefore has a hybrid architecture in which the sensor and readout electronics are manufactured and processed separately.

The CdTe sensor (ACRORAD Co., Ltd., Japan) is a 650 μm thick Schottky type diode with electrons collection on the pixels. It is characterized by a very low leakage current at 400÷500 V bias-voltage. The pixels are arranged on a hexagonal matrix with 60 μm horizontal pitch and 51.96 μm vertical pitch.
Table 1 summarizes the main characteristics of the CdTe sensor.

| Atomic numbers | 48, 52 |
| --- | --- |
| Effective atomic number | 50 |
| Density (g/cm$^3$) | 5.85 |
| Band energy (eV) | 1.5 |
| Dielectric constant | 11 |
| Ionizing energy (eV) | 4.43 |
| Resistivity (Ωcm) | $10^9$ |
| Electron mobility $\mu_e$ (cm$^2$/Vs) | 1100 |
| Electrons mean lifetime $\tau_e$ (s) | $3\times10^{-6}$ |
| Hole mobility $\mu_h$ (cm$^2$/Vs) | 100 |
| Hole mean lifetime $\tau_h$ (s) | $2\times10^{-6}$ |
| $(\mu\tau)_e$ (cm$^2$/V) | $3.3\times10^{-3}$ |
| $(\mu\tau)_h$ (cm$^2$/V) | $2\times10^{-4}$ |

Table 1 -The CdTe main characteristics

The CMOS ASIC has an active area of 30.7 × 24.8 mm², organized as a matrix of 512×476 pixels that match an identical pixel arrangement on the CdTe crystal. The chip integrates more than 350 million transistors. Each pixel incorporates a hexagonal electrode (top metal layer) connected to a charge amplifier that feeds two discriminators and two 15-bit counters/registers. To reduce unavoidable variations of the DC level between the pixels, a self-calibration circuit has been implemented in each pixel. In this way a single global threshold per discriminator can be applied to the entire matrix. During data acquisition, the shift registers are clocked by the trigger generated by their respective discriminators. After initialization, the number of events or clock periods recorded is uniquely linked to the register contents, with a maximum of $2^{15}$ counts per counter. In readout mode, the registers of several columns of pixels are serialized and their content pushed out of the circuit under the control of an external clock signal.

Two different data acquisition modes are selectable: 2 *colors* reading (2 thresholds, 2 counters) or, alternatively, single-threshold continuous reading (dead-time free mode, i.e. counting in one counter while reading the other one).

The color mode and the very high spatial resolution (50 μm) allow obtaining with the PIXIRAD sensors the optimal ratio between image quality and absorbed dose.

Table 2 summarizes the main characteristics of the CMOS ASIC.

| **Pixel characteristics** | |
|---|---|
| Shaped pulse duration (at the base) | 1 µs (adjustable) |
| Linear range | > 3000 electrons |
| Saturation level | > 6000 electrons |
| Equivalent noise (ENC) | 50 electrons (rms) |
| Residual offset after auto-calibration | ± 30 electrons |
| Maximum number of counts before reading | 32768 |
| Input signal | positive or negative |
| Possibility to disable, swap, by-pass pixel | user-selectable |
| **Pixel reading** | |
| Serialization of columns for best readout time | 16, 32, 64, 128 |
| Max readout clock frequency | 200 MHz |
| Readout time for 32 data outputs = 16 columns serialized (16 columns × 15 bits × 5 ns) | < 0.6 ms |
| Readout time for 16 data outputs = 32 columns serialized | < 1.2 ms |
| Readout time for 8 data outputs = 64 columns serialized | < 2.3 ms |

Table 2 - The CMOS ASIC characteristics

## 3. Images and performance

In depth testing of several configurations of the PIXIRAD imaging system has been performed. Modules of one, two, four and eight PIXIRAD tile units have been assembled with almost zero dead space between blocks, currently a 2 pixels column to be reduced to 1.

The modules are able to deliver extremely clear and highly detailed X-ray images for medical, biological, industrial and scientific applications in the energy range 1-100 keV. Images are obtained at very high count rate (> 30 GHz for a 4 tiles module).

Table 3 lists the main specs of PIXIRAD-1, a single unit system with 250K pixels, 500K counters and 3×2.5 cm² active area.

### 3.1 Chromatic photon counting

Thanks to the capability of the PIXIRAD modules to select the energy of the radiographic beam, radiological chromatic images with increased information content can be obtained, in real time and single exposure.

Fig.1 shows an example of chromatic photon counting with the single unit module, PIXIRAD-1.

| PIXIRAD-1 specs: a single unit Schottky type CdTe diode (650 μm thick, 30.9×25.0 mm$^2$) | |
|---|---|
| ASIC+CdTe base block | 512×476 pixels |
| Global active area | 31×25 mm$^2$ |
| Total number of pixels | 243712 |
| Total number of counters | 487424 |
| Pixel size | 60 μm hexagonal arrangement |
| Pixel density | 323 pixels/mm$^2$ |
| Pixel rate capability | 10$^6$ counts/pixel/s (dead-time corrected) |
| Global rate capability | 2.4×10$^{11}$ counts/s |
| Pixel dead-time | 300 ns |
| Position resolution | 11 line pair/mm at 50% MTF |
| Energy range | 1-100 keV |
| Detection efficiency @10 keV. 50 keV | 100%, 98% |
| Counters depth | 15 bits |
| Readout time at 50 MHz clock | 5 ms |
| Frame rate | 200 readouts/s (400 images/s) |
| Minimum applicable global threshold | 200 electrons |
| Sensor bias voltage | 200÷ 400 V |
| Leakage current density | 5 nA/cm$^2$ at 400 V, -20 °C |
| Typical number of defective pixels | Less than 1% |
| Number of independent thresholds (*colors*) | 2 set of two (swappable in real time) |
| Module size (W×L×H) | 14×14×7 cm$^3$ |
| Module weight | < 2 Kg |
| Module power consumption | 60 Watts (typical) |
| Module cooling | Liquid or forced air |
| Module operating temperature | +40 -40 °C |

Table 3 - Main characteristics of the PIXIRAD-1unit

The images of a small dry animal are obtained by simultaneously counting the X-ray photons with a low energy threshold (the LOW counter contains all photons), Fig. 1a, and a higher threshold, Fig. 1b (the HIGH counter contains the high energy photons). Fig. 1c) shows the same image generated by the low energy photons only. The image is obtained by subtracting the two previous pictures one from the other. From a single exposure, three *color* images are produced.

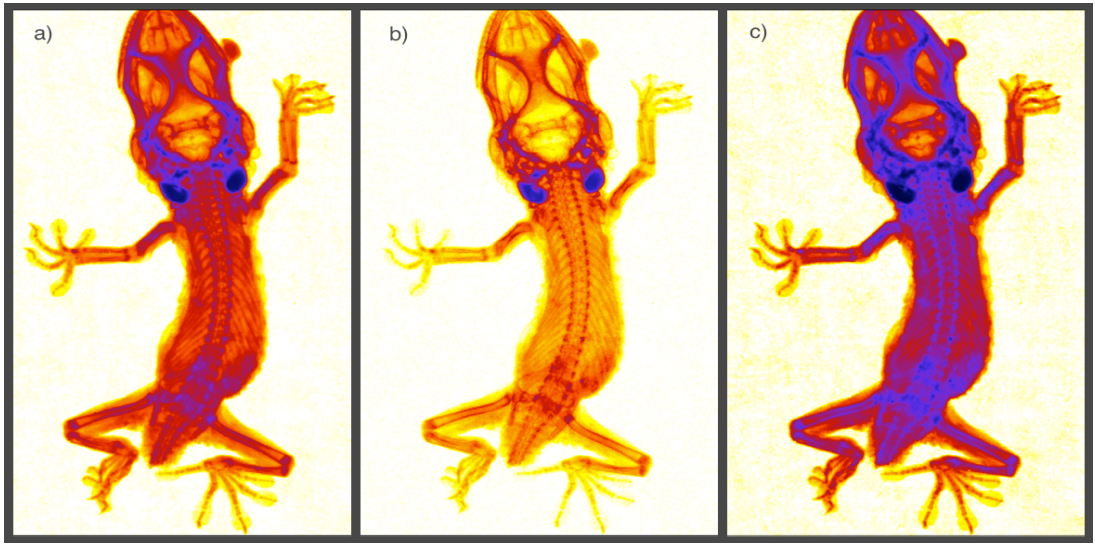

Fig. 1 - Images of a small dry animal obtained by simultaneously counting the X-ray photons with: a) a low energy threshold for counter 1 (all photons); b) a high energy threshold for counter 2 (high energy photons); c) subtracting each pixel content of image b) from the corresponding pixel of image a); the result is a low energy photons image.

**3.2 Low energy sensitivity**

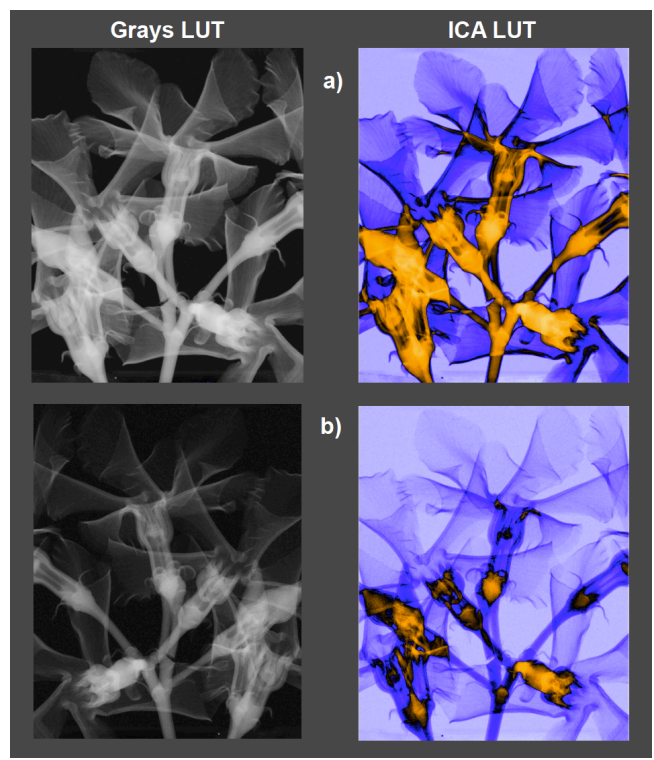

Fig. 2 - Images of a very low contrast object taken: a) at a global threshold of 200 electrons (1 keV energy, LOW counter), and b) at 1200 electrons (6 keV energy, HIGH counter).

The sensitivity of the PIXIRAD sensors to image low energy X-ray photons is shown in Fig.2. Here, images of a very low contrast object are taken at a global threshold of 200 electrons (1 keV energy, Fig. 2a) for counter 1, and 1200 electrons (corresponding to 6 keV energy, Fig. 2b) for counter 2.

For a better visualization of the differences in contrast between the images at 1 and 6 keV, two different Look-Up Tables (Grays and LUT) are used.

## 3.3 High energy sensitivity

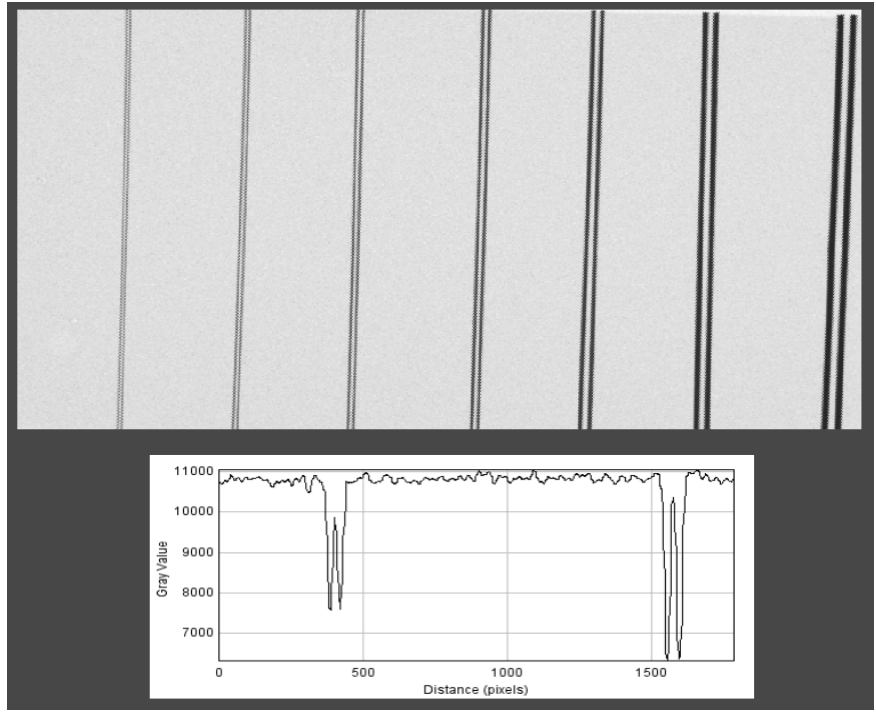

Fig. 3 - Image of a resolution phantom taken with a W anode X-ray tube set at 90 kVp. The plot is a profile across the last two couples of lines on the left. The closest lines are 50 μm wide 50 μm separation.

The image of a resolution phantom taken with a tungsten anode X-ray tube set at 90 kVp with 2 mm Aluminum filter, is shown in Fig. 3. The lines are metal strips embedded in a 5 mm thick FR4substrate (vetronite). Also at this high energy, the resolving power of the imaging system is extremely high, being the two closest lines 50 μm wide with 50 μm separation as it is shown by the profile plot taken across the last two couples of lines on the left side. No geometrical magnification has been used in the experimental set-up.

## 3.4 Spatial resolution

To measure the spatial resolution a Hüttner type spatial frequency grating (see Fig. 4a) has been used. The phantom is a rectangular thick lead foil embedded in glass with groups of bars, whose highest density is 10 line pairs per mm (lp/mm). A bar pattern may be considered resolved if the bars can be perceived with some discernible spacing between them. Fig. 4b shows a profile plot across the last four groups of bars including those at 10 lp/mm. At 50% MTF the resolution is 11 lp/mm.

## 4. The eight unit system PIXIRAD-8

The biggest in size PIXIRAD module that has been assembled to date, it is a 8 tiles unit, 25×2.5 cm² active area. The system has 2M pixels and 4M counters.

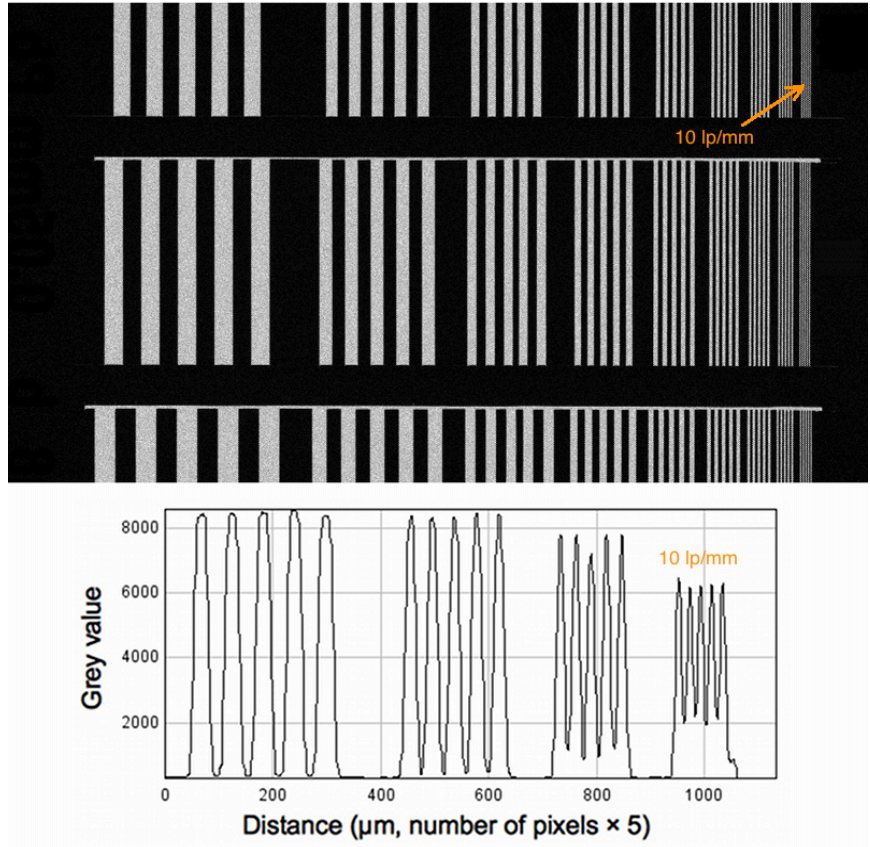

Fig. 4 – Image of a Hüttner test object for spatial resolution measurement

A single shot X-ray image of a man watch with its leather bracelet taken with PIXIRAD-8 is shown in Fig. 5. The watch is made of plastic and metal parts. The top image better visualizes low absorbing materials (plastic, leather), the bottom image better visualize the metal parts.

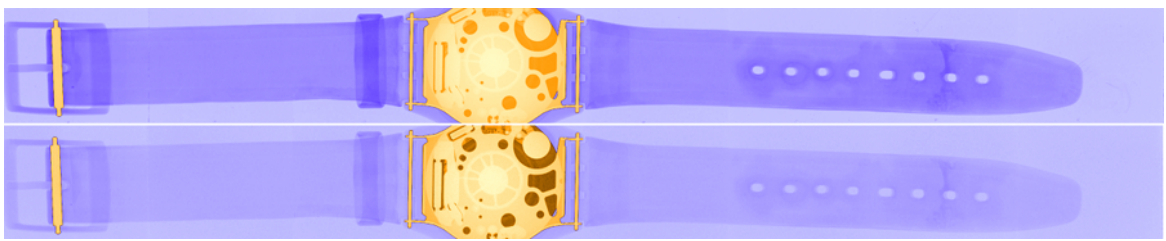

Fig. 5 – A single shot X-ray image of a man watch with its leather bracelet taken with PIXIRAD-8

## 5. Conclusions

The PIXIRAD X-ray sensor is an innovative, high quality, chromatic imaging system. It is intrinsically digital and noise free, due to the photon counting technology.
Its main characteristics are:
- optimal values of contrast and spatial resolution;
- high frame rate (~100 frame/sec);
- capability to separate the image in various color components depending on the incident radiation energy;
- capability to operate in dead-time free mode (reading one counter while taking data in the other one).

Starting from the base block, a complete imaging system can be obtained by coupling a number of PIXIRAD blocks along one direction.
An advanced system for Digital Mammography realized with the PIXIRAD-8 unit and operating in slot-scanning imaging mode, is under development by STM Electronics (Verona, Italy) in collaboration with PIXIRAD in the framework of an R&D project supported by Regione Friuli-Venezia Giulia (Italy).
The Digital Mammography is one of the most demanding X-ray imaging applications characterized by very fine position resolution, high sensitivity and DQE.
Next implementation will be the realization of a base block buttable along both X and Y directions to operate the system in full-field imaging mode.
The presented X-ray imaging system is the technological platform of PIXIRAD Imaging Counters s.r.l., a recently constituted INFN spin-off company (see http://pixirad.pi.infn.it/).